\documentclass[journal]{IEEEtran}
\pdfoutput=1
\usepackage{amsmath}
\usepackage{amssymb}
\usepackage{mathtools, cuted}
\usepackage{amsmath}
\usepackage{algpseudocode}
\usepackage{algorithm}
\usepackage{lipsum}
\usepackage{mathtools}
\usepackage{cuted}
\usepackage{amsmath}
\usepackage{amssymb}

\usepackage{graphicx}
\usepackage{stfloats,lipsum} 
\usepackage{epstopdf}
\usepackage[utf8]{inputenc}

\usepackage[pdftex]{hyperref}

\ifCLASSINFOpdf
\else
\fi

\begin{document}

\title{Dynamic Power Balancing Algorithm for Single-Phase Energy Storage Systems in LV Distribution Network with Unbalanced PV Systems Distribution}

\author{Watcharakorn Pinthurat,~\IEEEmembership{Student Member,~IEEE,} Branislav Hredzak,~\IEEEmembership{Senior Member,~IEEE,}

\thanks{Watcharakorn Pinthurat and Branislav Hredzak are with the school of Electrical Engineering and Telecommunications, the University of New South Wales, Sydney, NSW 2052, Australia, (email: w.pinthurat@student.unsw.edu.au; b.hredzak@unsw.edu.au).}
\thanks{}}

\markboth{}
{}

\maketitle
\begin{abstract}
Unbalanced power, due to high penetration of single-phase PV rooftops into a four-wire multi-grounded LV distribution system, can result in significant rise in the neutral current and neutral voltage. This preprint proposes a distributed clustering algorithm for dynamic power balancing, using single-phase battery storage systems distributed in the LV distribution system, in order to reduce the neutral current and neutral voltage rise. The distributed clustering algorithm aggregates households connected to the same phase into clusters. Within each cluster, another distributed clustering algorithm is applied to calculate the total grid power exchanged but the corresponding phase. Then, the dynamic power balancing control is applied to balance the powers at the bus, based on battery storage systems' charge/discharge constraints, power minimization  and willingness of the households to participate in the power balancing control.    
\end{abstract}


\IEEEpeerreviewmaketitle

\section{Introduction}

\IEEEPARstart{H}{igh} penetration of single-phase PV sources  into a low voltage (LV) distribution system has been increasing,  leading to unbalanced power among phases due to unbalanced allocation of the PV sources \cite{hossain2018multifunctional}. The unbalanced powers among the three phases can result in large neutral current and high neutral-to-ground voltage (NGV) \cite{balda1997measurements}. Large NGV can have negative effects on sensitive loads such as electronic devices and computing equipment because of the common-mode noise effect. NGV of  computing equipment should be maintained at less than $0.5$V as specified by manufacturers in \cite{gruzs1990survey}. Also, the high NGV can adversely affect  human beings as well as farm animals as reported in \cite{zipse2003hazardous} and \cite{lefcourt1991effects} respectively. Therefore, the neutral current and the NGV rise should be limited within predefined values to ensure safety and reliability in the LV distribution system.  

Several traditional methods to mitigate the neutral current and NGV rise by resizing neutral conductor, improving grounding and installing a passive harmonic filter were proposed in \cite{lefcourt1991effects}, \cite{surbrook1986stray}, \cite{zhu1998phase} and \cite{hsu1993transformer} . However, owing to high variation of single-phase PV units distributed in LV distribution system, power balancing control between loads and PV sources is difficult, and hence the traditional methods cannot properly solve the issues caused by the unbalanced powers. This is because the traditional methods are considered to be static \cite{alam2015community}. 
\begin{figure}[!t]
   \centering
    \includegraphics[width=85mm]{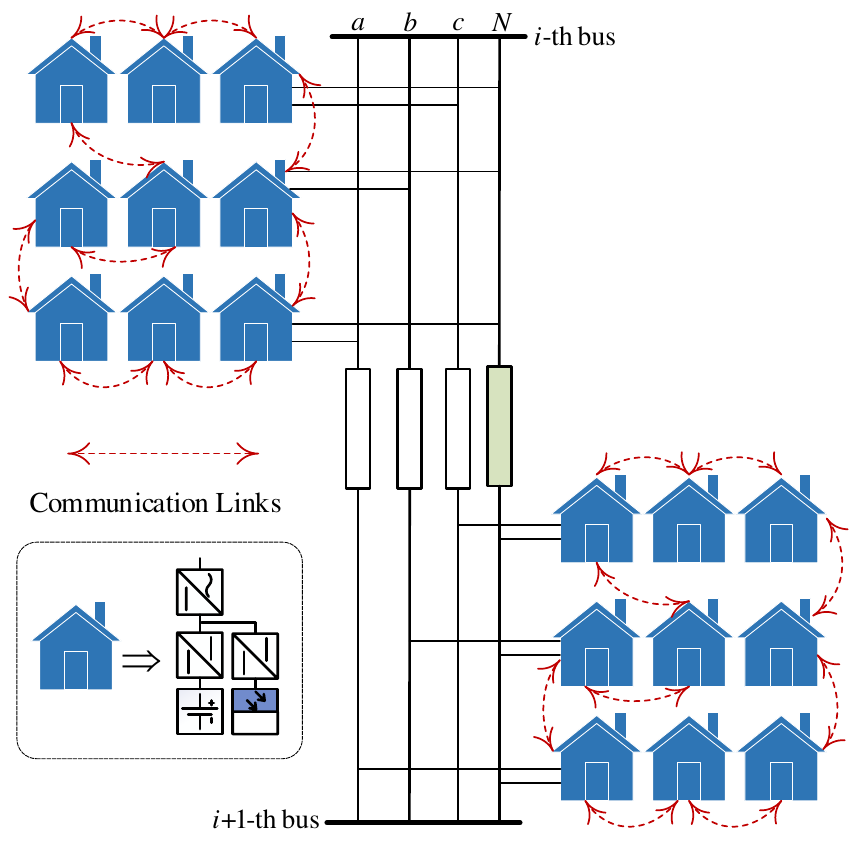}
    \caption{Schematic diagram of the proposed algorithm for balancing power in the LV distribution network. It is assumed that each household has a single-phase PV source and an ES device connected in parallel to a VSI, and has a local load. All households are connected via a sparse bidirectional communication network.}
    \label{LVSystem}
\end{figure}

There were various mitigation strategies proposed in literature to overcome the before mentioned issues. In \cite{alam2015community}, community energy storage (ES) was employed to mitigate NGV rise in four-wire multi-grounded LV distribution system. Although the neutral current and NGV rise were regulated within acceptable limits, this control strategy required DC common bus along the feeders to provide required active power to households during power balancing operation. Moreover, centralized communication links were used for households to communicate with the central energy storage. The main drawback of the centralized control is a single point of failure \cite{hoang2018accurate}. In \cite{alam2015alleviation}, the same authors addressed the neutral current and NGV rise by using single-phase distributed energy storage devices and distributed communication network. However, only one household in each phase was considered in a given bus. In reality, there are many households connecting to the same phase of the bus, as illustrated in Fig. \ref{LVSystem}. Therefore, different control approaches for phase power balancing in three-phase four-wire LV distribution system are required. In this preprint, a clustering based approach is proposed.
Clustering divides data into groups based on similar properties \cite{han2011data}. Several centralized clustering algorithms were proposed in literature, such as k-means, hierarchical clustering, self-organization map and expectation maximization clustering algorithm in \cite{yu2018two}, \cite{geva1999hierarchical}, \cite{flanagan1996self} and \cite{aci2011k} respectively. Distributed algorithms were also studied for distributed data environment \cite{visalakshi2009distributed}. However, a centralized network was still required for the distributed algorithms. This is because data storage from the distributed algorithms needs to communicate with a main site for clustering operation \cite{hai2012survey}. Besides, both centralized and distributed based clustering algorithms rely on  data collected from the previous time instant, not on real-time states. Hence, the above-mentioned algorithms may be inadequate for clustering households using the distributed communication network in LV distribution power system. 

Inspired by the above discussion, this preprint presents a distributed clustering algorithm for dynamic power balancing using distributed single-phase energy storage systems in order to mitigate the neutral current and neutral-to-ground voltage rise caused by unbalanced loads and PV sources in LV distribution system. The distributed clustering algorithm runs in real-time using dynamic states of multi-agent systems. The multi-agent based systems communicate via a sparse distributed communication network. The main contributions of this preprint are:
\begin{enumerate}
    \item  Distributed clustering algorithm to group households connected to the same phase into clusters is proposed. Each phase cluster has information about the real-time grid power exchanged by the other two phase clusters. 
    \item Dynamic power balancing control strategy, based on ESs' charge/discharge constraints, power minimization and willingness of the households to participate in the power balancing control, is combined with the distributed clustering algorithm  to alleviate the neutral current and neutral voltage rise in the network.
\end{enumerate}

The rest of the preprint is organized as follows. Section \ref{threephase} presents the active power flow in three-phase system. Section \ref{proposedalgorithm} describes the distributed clustering algorithms. Application of the proposed algorithm for power balancing control is discussed in Section \ref{application}. Finally, Section \ref{conclud} concludes the preprint. 

\section{Power Flow in Three-Phase System} \label{threephase}
\subsection{Neutral Current and Neutral Voltage Rise}
Neutral current is produced by unbalanced loads and PV units in a three-phase four-wire multi-grounded LV distribution system. A typical LV feeder including several households is shown in Fig. \ref{LVSystem}. There are a local load, an ES system and a PV unit in each household. The total neutral current $I_{N}$, produced by the imbalanced power at any given bus, is defined as, 
\begin{align}
    I_{N} &= -(I_{net}^{a} + I_{net}^{b} + I_{net}^{c}) \nonumber \\
    I_{net}^{\phi} &= \frac{(P_{l}^{\phi} + jQ_{l}^{\phi})^{*} - (P_{pv}^{\phi} + jQ_{pv}^{\phi})^{*}}{(V_{re}^{\phi} + jV_{im}^{\phi})^{*} - (V_{re}^{n} + jV_{im}^{n})^{*}},
\end{align}
where $*$ represents the complex conjugate; $P_{l}^{i\phi}$ and $P_{pv}^{i\phi}$ are the active powers of the local load and PV output power at the $\phi$-th phase; $Q_{l}^{i\phi}$ and $Q_{pv}^{i\phi}$ are the reactive powers of the local load and PV unit at the $\phi$-th phase; $V_{re}^{\phi}$ and $V_{im}^{\phi}$ are the real and imaginary parts of the phase voltage; $V_{re}^{n}$ and $V_{im}^{n}$ are the real and imaginary parts of the neutral voltage. 

To investigate the neutral current and its effects, the current unbalance factor (CUF) is widely employed \cite{kolagar2012effects}, calculated as, 
\begin{align}
    CUF &= \frac{\sqrt{|I_{ns}|^{2} + |I_{zs}|^{2}}}{|I_{ps}|} \times 100\%, \nonumber \\
     \begin{bmatrix}
        I_{ps}\\
        I_{ns} \\
        I_{zs}
        \end{bmatrix} &= \frac{1}{3} \begin{bmatrix}
                                        1 & a & a^{2}   \\
                                        1 & a^{2} & a   \\
                                        1 & 1 & 1
                                    \end{bmatrix} 
                                    \begin{bmatrix}
                                        I_{a} \\
                                        I_{b} \\
                                        I_{c} 
                                    \end{bmatrix}, a = e^{j(\frac{2\pi}{3})},
                                    \end{align} 
where $I_{ps}$, $I_{ns}$ amd $I_{zs}$ denote the positive, negative and zero sequences of current respectively. 
\subsection{Grid Power Exchange in Three-Phase Power System}
In a balanced three-phase system, for Y-connection and positive phase sequence, phase voltages are expressed as,
\begin{align}
\label{phasevol}
    v_{an}(t) &= V_{m}\sin (\omega t) \nonumber \\
    v_{bn}(t) &= V_{m}\sin (\omega t - 120^{\circ}) \nonumber \\
    v_{cn}(t) &= V_{m}\sin (\omega t + 120^{\circ}),
\end{align}
where $v_{an}$, $v_{bn}$ and $v_{cn}$ are the line-to-neutral voltages of phases $a$, $b$ and $c$ at any bus respectively, with a $120^\circ$ phase shift from the reference, and $V_{m}$ is the voltage magnitude. Note that phase $a$ is chosen to be the reference ($0^\circ$).

It is assumed that the power to be exchanged with the grid ($P_{g}^{i\phi}$) at each phase and the $i$-th bus is equal to the reference power ($P_{g}^{\text{ref-}i}$). The amount of phase power to be charged or discharged by the ES systems for the dynamic power balancing control can be calculated as,
\begin{equation}
\label{powerbalancing}
    P_{b}^{i\phi} = P_{g}^{\text{ref-}i} - P_{g}^{i\phi} = P_{g}^{\text{ref-}i} - (P_{l}^{i\phi} - P_{pv}^{i\phi}), \phi\in \{a,b,c\},
\end{equation}
where $P_{b}$ is the charged or discharged power by the distributed ES devices ($+$ve indicates the discharging mode while $-$ve indicates the charging mode), $P_{l}$ is the active load power, and $P_{pv}$ is the PV output power.

After the power mode $P_{b}^{i\phi}$ of the ES system is obtained from (\ref{powerbalancing}), the charging/discharging current of the ES device at the $\phi$-phases and the $i$-th bus can be defined as, 
\begin{equation}
\label{batterycurrent}
    I_{b}^{i\phi} = \frac{P_{b}^{i\phi}}{V_{b}^{i\phi}} = \frac{P_{b}^{i\phi}}{f(SoC_{b}^{i\phi}, C_{b}^{i\phi}, \Theta_{b}^{i\phi})}, \phi \in \{a,b,c\}, 
\end{equation}
where $I_{b}^{i\phi}$ is the charging/discharging current of the ES system, which will be used as a reference signal for the power balancing control, $V_{b}^{i\phi}$ is the output voltage of battery, depending on the battery state of charge (SoC), $C_{b}^{i\phi}$ is the battery capacity, and $\Theta_{b}^{i\phi}$ is a set of battery modelling parameters adopted from \cite{kim2011hybrid}.

\section{Dynamic Distributed Clustering Algorithm} \label{proposedalgorithm}
In this section, the distributed clustering algorithm for multi-agent based systems is introduced. It is defined that the algorithm will cluster $N$ agent into $M$ group based on pre-selected feature states. Within a cluster, the estimations of average states can also be accessed by each agent via the network. The algorithms was originally proposed in \cite{zhang2019novel} .
\subsection{Distributed Communication Graph}
A sparse graph $\mathcal{G}$($\mathcal{V}$,$\mathcal{E}$) represents distributed communication links among neighbours, where $\mathcal{V} = \{1,...,N \}$ and $\mathcal{E}$ denote the nodes (agents) and edges respectively \cite{lewis2013cooperative}. The node, $\mathcal{E}$, has elements $(i, j)$, in which $(i,j) \in \mathcal{E}$ if node $i$ can communicate with node $j$ via a communication link. The neighbours of the node $i$ are denoted as $\mathcal{N}_{i}$. Node $j$ is said to be a neighbour of node $i$ if $(i,j) \in \mathcal{E}$. The adjacent matrix of the communication graph is expressed by,
\begin{equation}
   \mathcal{A} = [a_{ij}] \in \mathbb{R}^{N\times N}, a_{ij} = \begin{cases}
      \alpha, (i,j) \in \mathcal{E} \\
      0, \text{otherwise}
    \end{cases},
\end{equation}
where $\alpha$ denotes the coupling gain. 

The graph Laplacian matrix is defined as,
\begin{equation}
    L = \mathcal{D} - \mathcal{A},
\end{equation}
where $\mathcal{D} =$ $diag$$\{d_{i}\}$, and the in-degree of the graph is represented as $d_{i}$ $=$ $\sum_{j=1}^{N}$$a_{ij}$.

In the following subsections, the distributed clustering algorithm, employing the distributed communication graph, is presented. 

\subsection{Real-time Distributed Clustering Algorithm}
In this subsection, the distributed clustering algorithm is introduced. The feature states of the $i$-th agent are denoted by $x_{i}$ $\in \mathfrak{R}^{n}$. These feature states are used to cluster $N$ agents into $M$ clusters in real-time. The average estimations of the states $x_{i}$ in cluster $j$ are defined as $\Bar{x}_{i}^{cstrj}$, in which $i = 1,...,N$ and $j = 1,...,M$. In all clusters, the latest average states $\Bar{x}_{i}^{cstrj}$ are dynamically accessed by each $i$-th agent via the distributed communication network.

Next, the algorithm is described. The initial average estimations of $\Bar{x}_{i}^{cstrj}$ are  selected arbitrarily. Then, the current state $x_{i}$ is compared to all current estimations of the $\Bar{x}_{i}^{cstrj}$, where $j = 1,...,M$ to decide to which cluster the $i$-th agent belongs to. The $i$-th agent belongs to the $k$-th cluster if its estimated state, $\Bar{x}_{i}^{cstrk}$, has the smallest distance from its current state $x_{i}$, as determined by (\ref{shortdistance}). After all clusters are obtained, for each $i$-th agent, average state estimation of the $k$-th cluster is compared with its current state $x_{i}$ to decide to which cluster it belongs to and the estimations of the $k$-th cluster agents are sent to the communication network, as defined by (\ref{two}). On the other hand, the state estimates of the neighbours, which are not members of the $k$-th cluster, $j \in \{1,...,N\}$, $j \neq k$, are passed through, as depicted in (\ref{three}) \cite{zhang2019novel}.
\begin{align}
    \dot{\Bar{x}}_{i}^{cstrk} &= \dot{x}_{i} + \sum_{l\in \mathcal{N}_{i}}a_{il}(\Bar{x}_{l}^{cstrk} - \Bar{x}_{l}^{cstrk}), \label{two}\\
    \dot{\Bar{x}}_{i}^{cstrj} &= \frac{1}{| \mathcal{N}_{i} |}\sum_{l \in \mathcal{N}_{i}}\dot{\Bar{x}}_{l}^{cstrj} + \sum_{l \in \mathcal{N}_{i}}a_{il}(\Bar{x}_{l}^{cstrj} - \Bar{x}_{i}^{cstrj}), j \neq k, \label{three}
\end{align}
where $|\mathcal{N}_{i}|$ is the number of neighbours of the $i$-th agent.

\begin{figure}[t!]
    \centering
    \includegraphics[width=85mm]{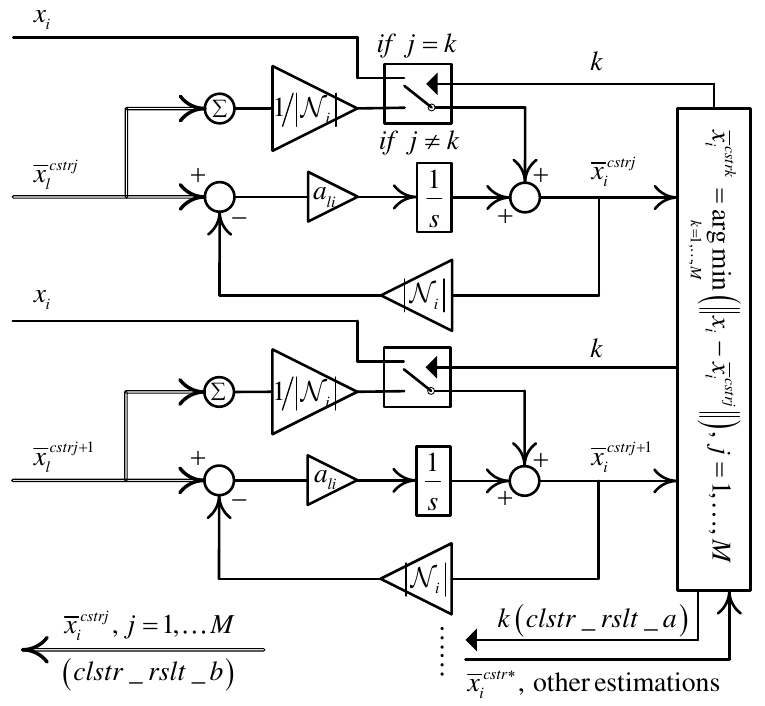}
    \caption{Real-time distributed clustering algorithm for the $i$-th agent. The distributed communication is represented by the double line arrows.}
    \label{Clustering}
\end{figure}

As depicted in Fig. \ref{Clustering}, the following two results can always be accessed by the $i$-th agent: 
\begin{enumerate}
    \item $clstr\_rslt\_a$: information that belongs to the $k$-th cluster;
    \item $clstr\_rslt\_b$: the average estimations of the feature states $\Bar{x}_{i}^{cstrj}$ in the $j$-th cluster, $j = 1,...,M$.
\end{enumerate}

The dynamic clustering algorithm is summarized as $\textbf{Algorithm \ref{DistributedClustering}}$ and Fig. \ref{Clustering} illustrates the implementation of the proposed algorithm.  
\begin{algorithm}
  \caption{Online Distributed Clustering}
  \label{DistributedClustering}
  \begin{algorithmic}[1]
  \State design $\mathcal{G}$($\mathcal{V}$,$\mathcal{E}$) and $M$
  \State initialize $\Bar{x}_{i}^{cstrj}$, $j = 1,...,M$
  \For {each $i$-th agent at time $t$}
  \State measure $x_{i}$
  \State receive $\Bar{x}_{l}^{cstrj}$, $l \in \mathcal{N}_{i}$, $j = 1,...,M$ from neighbours
  \State $k$ $\gets$ label of the cluster with the smallest distance between $x_{i}$ and $\Bar{x}_{i}^{cstrj}$, $j = 1,...M$, where $k \in \{1,...,M \}$
  \If {$j=k$}
  \State Eq. (\ref{two})
  \Else
  \State Eq. (\ref{three})
  \EndIf
  \State send $\Bar{x}_{i}^{cstrj}$, $j = 1,..., M$ to neighbours
  \EndFor
  \State Each local agent $i$ implements the same algorithm.
  \end{algorithmic}
\end{algorithm}

\begin{figure*}[!t]
\normalsize
\begin{align}
\label{shortdistance}
   x_{i} \in \Bigg\{ {{k\text{-}\text{th cluster} \Bigg | \Bar{x}_{i}^{cstrk} = \operatorname*{arg\,min}_{\{ k \in {1,...,M}\}}} \Big(\Big\| x_{i} - \Bar{x}_{i}^{cstrj} \Big\|\Big), j =1,...,M} \Bigg\}.
\end{align}
\hrulefill
\vspace*{4pt}
\end{figure*}

\subsection{Clustering Utilization}
\begin{figure}[ht]
    \centering
    \includegraphics[width=85mm]{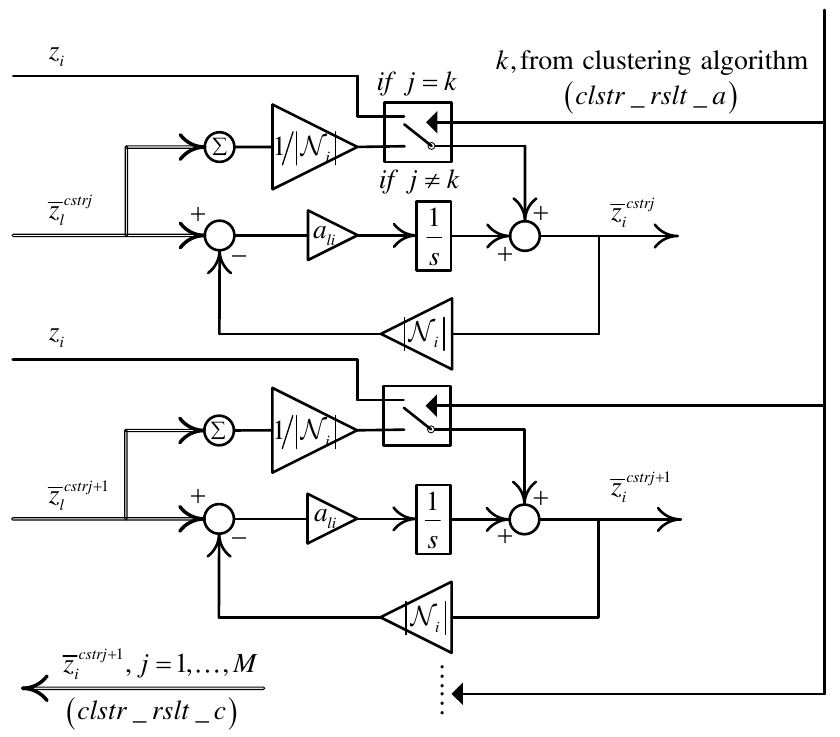}
    \caption{The estimates of the auxiliary state $z_{i}$ for the $i$-th agent. The distributed communication is represented by the double line arrows.}
    \label{Utilization}
\end{figure}

Within the clusters, the auxiliary states $z_{i} \in \mathfrak{R}^{m}$can be used for analysis. Therefore, the estimations of the auxiliary state, $z_{i}$ are defined as $\Bar{z}_{i}^{cstrj}$, where $i = 1,...,N$ and $j = 1,...,M$. Similar to the estimations of the average state, $\Bar{x}_{i}^{cstrj}$ from \textbf{Algorithm \ref{DistributedClustering}}, the $\Bar{z}_{i}^{cstrk}$ can be obtained using the result of $clstr\_rslt\_a$. Then, the same procedure as for the estimations of the feature state is followed \cite{zhang2019novel},
\begin{align}
    \dot{\Bar{z}}_{i}^{cstrk} &= \dot{z}_{i} + \sum_{l \in \mathcal{N}_{i}} a_{il}(\Bar{z}_{l}^{cstrk} - \Bar{z}_{i}^{cstrk}), \label{five} \\
    \dot{\Bar{z}}_{i}^{cstrj} &= \frac{1}{| \mathcal{N}_{i}|}\sum \dot{\Bar{z}}_{l}^{cstrj} + \sum_{l\in  \mathcal{N}_{i}} a_{il}(\Bar{z}_{l}^{cstrj} - \Bar{z}_{i}^{cstrj}), j \neq k. \label{six}
\end{align}
The result of $clstr\_rslt\_c$ is accessed by the $i$-th agent node via the distribution network as illustrated in Fig. \ref{Utilization}. 

\textbf{Algorithm \ref{utilized}} is obtained by including the average estimations of auxiliary states into \textbf{Algorithm \ref{DistributedClustering}}.
\begin{algorithm}
  \caption{Utilization of Auxiliary States within Cluster}
  \label{utilized}
  \begin{algorithmic}[1]
  \State design $\mathcal{G}$($\mathcal{V}$,$\mathcal{E}$) and $M$
  \State initialize $\Bar{x}_{i}^{cstrj}$, $j = 1,...,M$
  \For{each $i$-th agent at time $t$}
  \State measure $x_{i}$ and $z_{i}$
  \State receive $\Bar{x}_{l}^{cstrj}$, $l \in \mathcal{N}_{i}$, $j = 1,...,M$ from neighbours
  \State receive $\Bar{z}_{l}^{cstrj}$, $l \in \mathcal{N}_{i}$, $j = 1,...,M$ from neighbours
  \State $k$ $\gets$ label of the cluster with the smallest distance between $x_{i}$ and $\Bar{x}_{i}^{cstrj}$, $j = 1,...M$, where $k \in \{1,...,M \}$
  \If {$j=k$}
  \State Eq. (\ref{two}) and (\ref{five})
  \Else
  \State Eq. (\ref{three}) and (\ref{six})
  \EndIf
  \State send $\Bar{x}_{i}^{cstrj}$, $j = 1,..., M$ to neighbours
  \State send $\Bar{z}_{i}^{cstrj}$, $j = 1,..., M$ to neighbours
  \EndFor
  \end{algorithmic}
\end{algorithm}

\section{Application Example} \label{application}
In this section, an application of the distributed clustering algorithm to dynamic power balancing  for mitigating neutral current and neutral voltage rise caused by imbalance of PV sources and load demands is introduced. 

ES devices powers, for balancing operation in (\ref{powerbalancing}),  should be minimized to reduce required ES capacity. In the case that the power exchange with the grid in all phases has the same sign (either consumed or delivered powers), the reference power $P_{g}^{\text{ref-}i}$ at a given bus is obtained as,  
\begin{equation}
\label{Pgref}
    P_{g}^{\text{ref-}i} = \text{min}(|P_{g}^{ia}|, |P_{g}^{ib}|, |P_{g}^{ic}|).
\end{equation}

However, if the power exchange with the gird in all phases and at a given bus has different signs (both power consumption and delivery), (\ref{optimization}) is required to ensure the sum of power injected and consumed by ES systems in a given bus is minimized. 
\begin{align}
    &\text{minimize:} \sum_{\phi\in \{a,b,c\}} |P_{b}^{i\phi}|, \text{subject to}:  \label{optimization}\\
    &\begin{bmatrix}
        0\\
        0\\
        0
    \end{bmatrix} = \begin{bmatrix}
                        1 & -1 & 0\\
                        0 & 1 & -1\\
                        -1 & 0 & 1
                    \end{bmatrix}\begin{bmatrix}
                        P_{b}^{ia}\\
                        P_{b}^{ib}\\
                        P_{b}^{ic}
                    \end{bmatrix} + \begin{bmatrix}
                        P_{g}^{ia}-P_{g}^{ib}\\
                        P_{g}^{ib}-P_{g}^{ic}\\
                        P_{g}^{ia}-P_{g}^{ic}
                    \end{bmatrix}, \label{equality}\\
    &(P_{b}^{ia}\times P_{b}^{ib}) \geq 0, (P_{b}^{ib}\times P_{b}^{ic}) \geq 0, (P_{b}^{ia}\times P_{b}^{ic}) \geq 0. \label{inequality}
\end{align}

The equality constraint in (\ref{equality}) ensures equal power exchange with the grid for all  three phases while the correct decision made by ES systems (charging/discharging mode) is ensured by the inequality constraints in (\ref{inequality}) \cite{alam2015alleviation}.
\begin{figure}[!t]
    \centering
    \includegraphics[width=80mm]{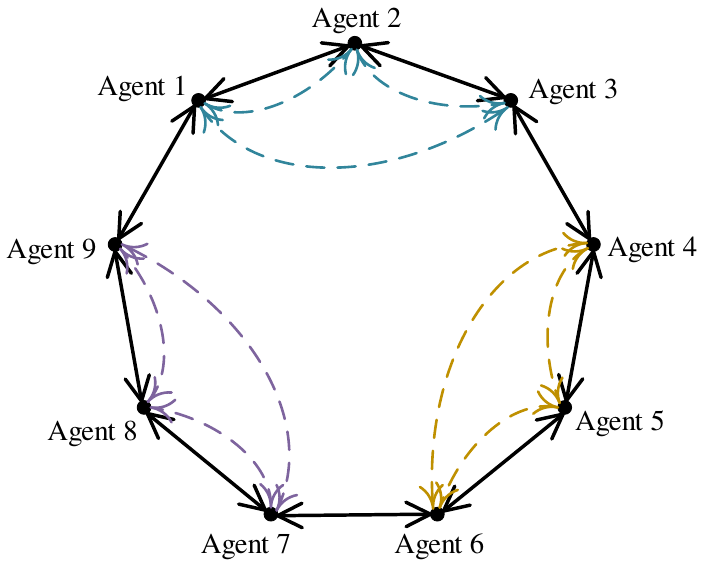}
    \caption{Illustration of the proposed clustering algorithm. Nine agents are clustered into three groups. Each agent represents a household at a bus. The black lines represent distributed communication network between agent nodes. The colored dashed lines indicate virtual communications with a cluster. One color represents one cluster.}
    \label{Agent}
\end{figure}
\begin{figure}[b!]
    \centering
    \includegraphics[width=85mm]{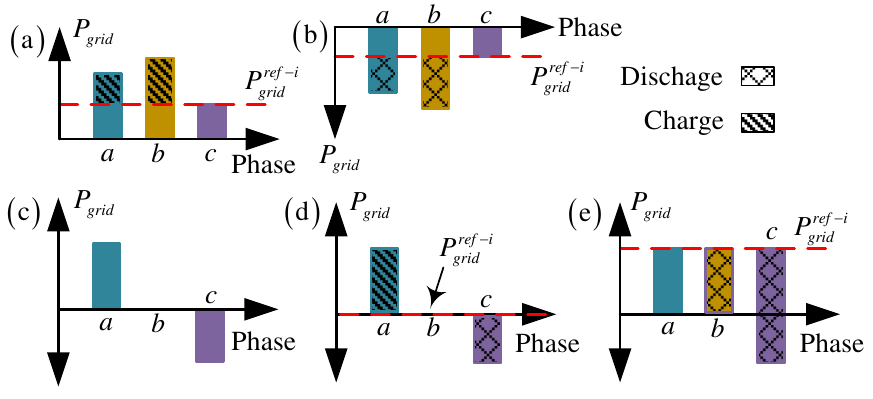}
    \caption{Different scenarios of the power exchange with the grid (a) $P_{g}$ $>$ $0$ (for all phases), (b) $P_{g}$ $<$ $0$ (for all phases), (c)  $P_{g}$  varies (both power delivery and consumption), and (d) both charging and discharging modes to achieve the  power balancing, and (e), only discharging mode for the power balancing.}
    \label{Scenarios}
\end{figure}
\subsection{Methodology}
Methodology for the distributed clustering algorithm with dynamic power balancing control can be described as follows: 

\begin{enumerate}
    \item In a given bus, the active powers among the three phases should be equally balanced, a sum of the phase currents is equal to zero ($I_{a}+I_{b}+I_{c}=0$). It is assumed that there are nine households connected to a bus, and three households to a phase, as shown in Fig. \ref{LVSystem}. 
    \item \textbf{Algorithm \ref{DistributedClustering}} is required to group the households at the same bus into three clusters (phases $a$, $b$ and $c$) by using the phase shifts in  (\ref{phasevol}) as the feature states. Virtual communication links are formed as shown in Fig. \ref{Agent}. A cluster of Agents $1$-$3$ represents the households belonging to phase $a$, while the other two household clusters belong to phases $b$ and $c$.  
    \item \textbf{Algorithm }\ref{utilized} is adopted for utilization of the feature states within the cluster. In this step, the grid power exchanges ($P_{g}^{i\phi}$) are calculated using  (\ref{powerbalancing}) and are subsequently sent to the other two clusters. (Note that each cluster has to be able to communicate with the neighbouring clusters.)
    \item Next, two possible decisions can be made by the power balancing control method. If the the grid power exchanges ($P_{g}^{i\phi}$) have  the same sign (either injected power or consumed power from the grid), ESs' powers ($P_{b}^{i\phi}$) for balancing operation are calculated using (\ref{Pgref}). However, if the grid power exchanges ($P_{g}^{i\phi}$) have opposite signs (both delivered and consumed powers), (\ref{optimization}) along with (\ref{equality}) and (\ref{inequality}) are used to calculate the ESs' powers ($P_{b}^{i\phi}$).  
    \item After ESs' powers ($P_{b}^{i\phi}$) are obtained, charge/discharge currents of the ES devices are calculated using  (\ref{batterycurrent}) to obtain the reference signals for the power control loop. In this step, $SoC_{max}$ and $SoC_{min}$ of the ES devices are taken into consideration to ensure that all ES devices operate within the limits.
    \item In a realistic situation, not all households may be willing to participate in the power balancing control even though they have sufficient power. In this case, the average state estimations in the clustering algorithm will not include the unwilling households in the power control operation. 
    \item Furthermore, SoC condition may be taken into consideration. Batteries having a certain SoC, eg close to $50\%$, are more suitable to participate in the power balancing control. The households having too high or too low SoCs may not be allowed to participate into power control even though they are willing to. 
\end{enumerate}
\subsection{Grid Power Exchange Scenarios} 
There are three possible scenarios of the grid power exchanges in the four-wire multi-grounded LV distribution system illustrated in Fig. \ref{Scenarios} :
\subsubsection{All three phases inject powers to the grid} In Fig. \ref{Scenarios}(a),  $P_{g}$ is greater than zero for all phases (inject powers to the grid). The algorithm finds the phase which injects the minimum power to the grid using (\ref{batterycurrent}) (phase $c$ in this example). Then,  the ES devices distributed in the phases $a$ and $b$  are charged to absorb the surplus powers and make the powers injected to the grid equal to that of the phase $c$. Hence, the active powers for all phases at a given bus are balanced.

\subsubsection{All three phases consume powers from the grid} $P_{g}$  is less than zero for all phases (powers consumed by local loads), as depicted in Fig. \ref{Scenarios}(b). The algorithm finds the phase with the minimum power (absolute value) consumed from the grid using (\ref{batterycurrent}) (phase $c$ in this example). Then,  the ES devices distributed in the phases $a$ and $b$  are discharged to make the powers consumed from the grid equal to that of the phase $c$. Hence, the active powers for all phases at a given bus are balanced.

\subsubsection{Phases inject/consume powers to/from the grid} In this case, the powers are both injected to the grid and consumed by loads from the grid  simultaneously, as shown in Fig. \ref{Scenarios}(c). Power balancing is achieved by both charging and discharging or only discharging the ES devices as shown in Figs. \ref{Scenarios}(d) and \ref{Scenarios}(e) respectively. In these cases, (\ref{optimization}) along with (\ref{equality}) and (\ref{inequality}) are required to find the total minimum power consumed or delivered by the ES devices to balance operation at a given bus.  
\begin{figure}[t!]
    \centering
    \includegraphics[width=85mm]{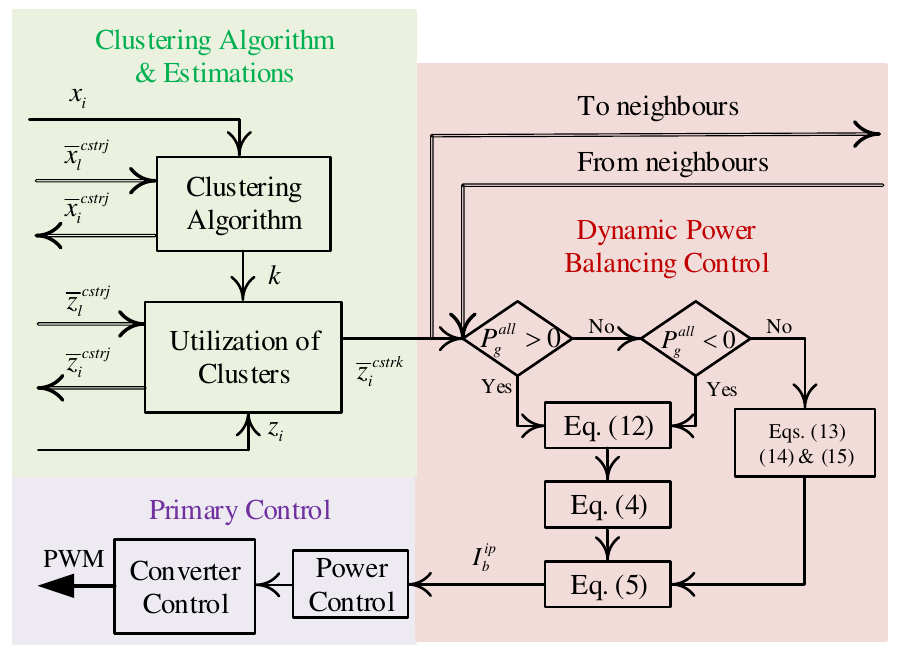}
    \caption{Control block diagram of the proposed control system for the $n$-th household.}
    \label{ControlDiagram}
\end{figure}

\subsection{Control System Implementation}
Fig. \ref{ControlDiagram} depicts the complete control system of the proposed control strategy for the $i$-th ES system.  
\begin{enumerate}
    \item Green block shows the distributed clustering algorithm and the auxiliary states estimations. In the clustering algorithm block,  the phase shift in (\ref{phasevol}) is the pre-selected feature state $x_{i}$. In this block, a virtual ES system  is formed at each phase at a given bus. Within the cluster, the obtained auxiliary average states $\Bar{z}_{i}^{cstrk}=$ $P_{g}^{i\phi}$ ($= P_{l}^{i\phi} - P_{pv}^{i\phi}$) are sent to the neighbours and  to the power balancing control.
    \item Red block represents the dynamic power balancing control. In this block, the grid power exchanges from the other two phases are received from the neighbours. Hence, the battery current in (\ref{batterycurrent}) for the power balancing can be obtained.
    \item Purple block indicates that the current from the dynamic power balancing control is the reference signal for the power control block at the primary level control. The converter controller generates the PWM signal for the converter to achieve the power balancing control. 
\end{enumerate}

\section{Conclusion} \label{conclud}
This preprint presented a distributed clustering algorithm with dynamic power balancing control  using single-phase distributed battery energy storage systems to alleviate the neutral current and neutral voltage rise produced by unbalanced PV allocations in four-wire multi-grounded LV distribution network. The distributed clustering algorithm was used to group the households at the same phase into the same cluster. Within the cluster, another distributed clustering algorithm was applied to calculate the total grid power exchanged by each phase. Then, the dynamic power balancing control was adopted to balance the powers among the phases at a bus, considering charge/discharge constraints, power minimization and willingness of the households to participate in the power balancing control.

\section*{Acknowledgment}

W. Pinthurat would like to gratefully thank for a scholarship from the Ministry of Higher Education, Science, Research and Innovation, Thailand.

\bibliographystyle{IEEEtran}
\bibliography{bibfile}

\end{document}